\documentclass[3p,twocolumn]{elsarticle}
\usepackage{amsmath}
\usepackage{amssymb}

\begin{document}

\title{High energy improved scalar quantum field theory\\ from noncommutative geometry without UV/IR-mixing}

\author{Alexander Schenkel}
\ead{aschenkel@physik.uni-wuerzburg.de}
\author{Christoph F.~Uhlemann}
\ead{uhlemann@physik.uni-wuerzburg.de}

\address{Institut f\"ur Theoretische Physik und Astrophysik\\
Universit\"at W\"urzburg\\
Am Hubland, 97074 W\"urzburg, Germany}


\begin{abstract}
We consider an interacting scalar quantum field theory on noncommutative Euclidean space. 
We implement a family of noncommutative deformations, which -- in contrast to the well known 
Moyal-Weyl deformation -- lead to a theory with modified kinetic term, while all local potentials are unaffected by the deformation. 
We show that our models, in particular, include 
propagators with anisotropic scaling $z=2$ in the ultraviolet (UV). 
For a $\Phi^4$-theory on our noncommutative space we obtain an improved UV behaviour 
at the one-loop level and the absence of UV/IR-mixing and of the Landau pole.
\end{abstract}
\begin{keyword}
Noncommutative geometry\sep star-products\sep quantum field theory.
\PACS 11.10.Nx\sep 02.40.Gh\sep 03.70.+k.
\end{keyword}

\maketitle

\section{Introduction}
Quantum field theories (QFTs) on noncommutative (NC) spacetimes are of recent interest, not only in mathematical physics, 
but also in particle phenomenology and cosmology. Historically, one of the main reasons for the introduction of NC spacetimes 
was the hope for a regularization of the divergences occurring in QFT,
leading to a better behaved theory. Indeed, as it has been shown by Grosse and Wulkenhaar \cite{Grosse} 
and later also by the group
around Rivasseau \cite{Rivasseau}, introducing the canonical NC $[x^\mu,x^\nu]=i\Theta^{\mu\nu}$, where 
$\Theta^{\mu\nu}$ is constant, and adding a harmonic oscillator potential to the action leads to an improvement of the 
quantum properties of the $\Phi^4$-theory, see also \cite{RGE} for related work. 
More precisely, the Grosse-Wulkenhaar model does not have a Landau pole, is
renormalizable to all orders in perturbation theory and can be seen as a good candidate for a rigorous interacting QFT in
four dimensions.  

Despite its mathematical beauty, this theory has two phenomenological drawbacks. Firstly, it is formulated on Euclidean space
and it is not obvious how to generalize it to the Minkowski spacetime. However, see \cite{Fischer}
 for a recent approach in this direction.
Secondly, the additional term introduced in the action in order to obtain a renormalizable theory leads to a strong 
infrared (IR) modification of the propagator, which will make it very hard to render theories of this kind
 compatible with experiments in particle and astrophysics. The 
phenomenologically problematic extra term in the action can be traced back to an infamous feature called UV/IR-mixing,
which is typically present in QFTs on the canonically deformed Euclidean space, but also on other NC
spaces such as $\kappa$-deformed space \cite{Grosse:2005iz} or in the continuum limit of fuzzy spaces, see e.g.~\cite{fuzzyUVIR}.
 The reason for the UV/IR-mixing  lies in the phase factors present in 
the modified Feynman rules, which are due to a NC potential $\Phi^{\star4}$, where $\star$ denotes
the Moyal-Weyl $\star$-product realizing the canonical NC space.

In this short letter we present a mechanism by which NC geometry can lead to an improvement of scalar QFT without
the second phenomenological drawback mentioned above, i.e.~we consider a deformed Euclidean space and find
neither UV/IR-mixing, nor an IR modified propagator in our theory. More precisely, we show that there is a NC deformation 
of the Euclidean space, which behaves just opposite to the Moyal-Weyl deformation 
in the sense that the kinetic term of the scalar field action gets deformed, while the potential terms remain commutative. 
We show that a special case of our deformation leads to a propagator with a $z=2$ anisotropic scaling in the UV, 
i.e.~the propagator denominator has the form $E^2 + {\bf{k}}^2 + \alpha^2\, {\bf{k}}^4+m^2$. 
Theories with anisotropic scaling ${\bf{k}}^{2z}$, with $z>1$, lead to an improved UV behaviour
 in perturbation theory, such that even gravity is expected to be perturbatively renormalizable for $z\geq3$ \cite{Horava:2009uw}. 
 We find that the one-loop  correction to the two-point function in our deformed $\Phi^4$-theory (i.e.~the tadpole diagram)
 diverges only with the square root of the momentum space cutoff and that the one-loop correction to the 
 four-point function is finite, thus avoiding the Landau pole.

The outline of this paper is as follows: In Section \ref{sec:deformation} we introduce the framework in which we construct
NC spaces. In Section \ref{sec:action} we deform the action of a scalar field consisting of a 
standard kinetic term and a local potential. We obtain a class of deformations, where the potential remains undeformed,
but the kinetic term is modified. We study some physical properties of our theory, including the renormalization of the propagator
and the four-point vertex at the one-loop level in Section \ref{sec:physics}. We conclude in Section \ref{sec:conclusion}.

\section{\label{sec:deformation}The noncommutative deformation}
Before going to our class of NC deformations, we briefly review the standard Moyal-Weyl deformation of the Euclidean
space. The Moyal-product of two smooth functions $f,h$ on $\mathbb{R}^{d}$ is given by
\begin{flalign}
f\star_M h := f\,e^{\frac{i\lambda}{2}\overleftarrow{\partial_\mu}\Theta^{\mu\nu}\overrightarrow{\partial_\nu}}\,h~,
\end{flalign}
where $\Theta^{\mu\nu}=-\Theta^{\nu\mu}$ is real and constant, 
leading to the commutation relations $[x^\mu\stackrel{\star_M}{,}x^\nu]:=x^\mu\star_M x^\nu -x^\nu\star_M x^\mu=i\lambda\Theta^{\mu\nu}$. We can generalize this $\star$-product 
by allowing instead of the partial derivatives more general vector fields $X_a=X_a^{\mu}(x)\partial_\mu$ 
satisfying $[X_a,X_b]=0$ for all $a,b$, where 
the vector field commutator is given by the Lie bracket. This leads to $\star$-products of RJS-type \cite{Reshetikhin:1990ep,Jambor:2004kc}
\begin{flalign}
\label{eqn:starprod}
f\star h := f\,e^{\frac{i\lambda}{2}\overleftarrow{X_a}\Theta^{ab}\overrightarrow{X_b}}\,h~,
\end{flalign}
where the vector fields are defined to act on functions via the Lie derivative and $\Theta^{ab}$ can be fixed to 
the canonical (Darboux) form without loss of generality
\begin{flalign}
\label{eqn:canonical}
 \Theta^{ab} = \begin{pmatrix}
         0 & 1 & 0 & 0 &\cdots \\
	-1 & 0 & 0 & 0 &\cdots\\
	 0 & 0 & 0 & 1 &\cdots\\
	 0 & 0 & -1 & 0 & \cdots \\
	 \vdots & \vdots & \vdots & \vdots & \ddots
        \end{pmatrix}~.
\end{flalign}
The associated commutation relations are given by $[x^\mu\stackrel{\star}{,}x^\nu]=i\lambda \Theta^{ab} X_a^\mu(x) X_b^\nu(x)+\mathcal O(\lambda^2)$.
Since we are interested in field theory, where also (exterior) derivatives of fields occur, 
we have to extend the $\star$-product to a deformed product between
functions and one-forms. This has been done by Wess {\it et al.}~\cite{Aschieri} and one simply finds 
for the $\star$-product of a function $f$ and a one-form $\omega$
\begin{flalign}
\omega\star f = \omega\,e^{\frac{i\lambda}{2}\overleftarrow{X_a}\Theta^{ab}\overrightarrow{X_b}}\,f~,
\end{flalign}
where the vector fields act on forms via the Lie derivative, as well. The $\star$-multiplication $f\star \omega$ is defined completely
analogous. This construction can be generalized to deformed wedge products of forms and, using the undeformed exterior
 differential $d$, we obtain a deformed differential calculus \cite{Aschieri}. Note that the deformed differential calculus is essential for
the construction of actions, since the derivatives appearing there imply that the action not only depends on the algebra of
functions but also on the differential calculus. 

Let us now specialize the $\star$-product (\ref{eqn:starprod}). We choose one direction, say $x^0$, of Euclidean space 
$(\mathbb{R}^{d},g)$, where $g=dx^\mu\otimes g_{\mu\nu}\, dx^\nu$ and $g_{\mu\nu}=\text{diag}(1,1,\dots,1)_{\mu\nu}$ 
is the metric. The orthogonal coordinates are denoted by $x^i$, $i=1,\dots,d-1$. Consider now vector fields $X_a$, $a=1,\dots, 2N$, 
defined by
\begin{subequations}
\label{eqn:vectorfields}
\begin{flalign}
& X_{2n-1}=T_{n}^{~i}\,\partial_i~,\\
& X_{2n}=\vartheta(x^0)\,T_{n}^{~i}\,\partial_i~,
\end{flalign}
\end{subequations}
where $T_n^i$ are constant and $\vartheta$ is a smooth function of $x^0$.
Note that for all $n$ $X_{2n-1}$ and $X_{2n}$ are parallel, i.e.~they have the the same $T_n^{~i}$. At first sight, one might expect
that this deformation is trivial, since every two smooth functions $f,h$ commute, $[f\stackrel{\star},h]=0$.
But this is not the case, since the $\star$-product acts nontrivially between functions and forms. In particular, we find
for the basis one-forms $dx^\mu$ and a smooth function $f$
\begin{subequations}
\label{eqn:starbasis}
\begin{flalign}
dx^0\star f &= dx^0\, f~,~\\
 dx^i\star f &= dx^i\,f -dx^0\, \frac{i\lambda}{2}\dot\vartheta(x^0)\,T_n^i\,T_n^j\partial_j f~,
\end{flalign}
\end{subequations}
where dot denotes the derivative with respect to $x^0$ and the sum over $n$ is understood. 
Note that higher order terms in $\lambda$ vanish in the $\star$-product (\ref{eqn:starbasis}), 
thus convergence is trivially given. 

\section{\label{sec:action}The deformed scalar field action }
We now construct an action for a real scalar field $\Phi$ deformed by (\ref{eqn:starprod}) and the choice of vector fields
 (\ref{eqn:vectorfields}). We follow the construction presented in \cite{Ohl:2009qe}, which makes extensive use of the deformed differential calculus.
 We present the following steps in a constructive way and refer to \cite{Ohl:2009qe} for a detailed introduction to the formalism.
 For deriving the action it is important to write the metric tensor in a basis with $\star$-products, 
 i.e.~$g=dx^\mu\otimes_\star g_{\star\mu\nu}\star dx^\nu$, where the coefficients $g_{\star\mu\nu}$ are 
 in general different from the metric coefficients in the standard basis. However, using our specific $\star$-product, 
 we find that $g_{\star\mu\nu}=g_{\mu\nu}$. An additional ingredient required to formulate the action is the volume form.
 On the Euclidean space it is given by $\text{vol}=dx^0\wedge dx^1\dots\wedge dx^{d-1}$. Note that all 
 wedge products in the volume form
 can be identically replaced by $\star$-wedge products in our deformation. 
 Furthermore, note that the Lie derivative of $\text{vol}$ along 
 any of the vector fields $X_a$  (\ref{eqn:vectorfields})  is zero, i.e.~$\mathcal{L}_{X_a}\text{vol}=0$ for all $a$. Following \cite{Ohl:2009qe}, 
 the deformed 
 scalar field action can be expressed in the basis including $\star$-products as
 \begin{flalign}
 S_\star =  \int \left(\frac{1}{2}(\partial_{\star\mu}\Phi)^\ast\star g_{\star}^{\mu\nu}\star\partial_{\star\nu}\Phi + V_\star[\Phi]\right)\star\text{vol}~,
 \end{flalign}
 where $g_\star^{\mu\nu}=g^{\mu\nu}=\text{diag}(1,1,\dots,1)^{\mu\nu}$ is the inverse metric, 
 $V_\star[\Phi]$ is a potential term including
 $\star$-products and $\partial_{\star\mu}$ are the deformed derivatives  obtained by comparing both sides of
  $dx^\mu\,\partial_\mu\Phi = dx^\mu \star(\partial_{\star\mu}\Phi)$. In our specific example, $\star$-products of scalar fields $\Phi$ reduce
  to the standard products and the $\star$-product acts trivially
  on the metric coefficients $g^{\mu\nu}$ and the volume form $\text{vol}$. 
  Using this and (graded) cyclicity, which allows us to remove one $\star$-product 
  under the integral,  we find that the only remaining deformation is due to the deformed derivatives $\partial_{\star\mu}$.
We obtain using (\ref{eqn:starbasis}) (sum over $n$ understood)
\begin{subequations}
\begin{flalign}
\partial_{\star 0} &= \partial_0+\frac{i\lambda}{2}\dot\vartheta(x^0)T_n^{~i}T_n^{~j}\partial_i\partial_j\Phi~,\\
\partial_{\star i} &= \partial_i~.
\end{flalign}
\end{subequations}
This leads to the deformed action
\begin{flalign}
\label{eqn:action}
S_\star=\int \left(\frac{\partial^\mu\Phi\partial_\mu\Phi}{2} + \frac{\lambda^{2}}{8}\dot\vartheta^2(\mathcal{T}\mathcal{T}\Phi)^2 +V[\Phi]\right)\,\text{vol}~,
\end{flalign}
where we defined $\mathcal{T}:=T_n^{i}\partial_i$ and the sum over $n$ is implicit.
Note that this is an exact result to all orders in $\lambda$ and {\it not} an expansion in $\lambda$, 
since the $\star$-products automatically terminate at $\mathcal O(\lambda)$.

\section{\label{sec:physics}Physical properties of the model}
\subsection{\label{par:z2}$z=2$ scalar field propagator}
The action of a scalar field with a $z=2$ anisotropic scaling can be obtained from (\ref{eqn:action}) by choosing $N=d-1$,
$T_n^{~i}=\delta_n^{i}$ and $\vartheta(x^0) = 2 \gamma\, x^0$ (the factor $2$ is for later convenience), since in this case we have 
\begin{flalign}
\mathcal{T}\mathcal{T}=\sum_{n=1}^{d-1} \delta_n^i\delta_n^j \partial_i\partial_j = \partial^i\partial_i~.
\end{flalign}
Including a quadratic potential $V[\Phi]=\frac{1}{2}m^2\Phi^2$ in (\ref{eqn:action}), the propagator 
denominator for this choice of deformation is given by
\begin{flalign}
\label{eqn:prop}
E^2+{\bf{k}}^2 +\alpha^2\,{\bf{k}}^4+m^2~,
\end{flalign}
where $\alpha^2=\lambda^2\,\gamma^2>0$ and $E/{\bf k}$ is the momentum associated to $x^0/{\bf x}$. 
Propagators of this form became of recent interest due to Ho\v rava's proposal of a power-counting renormalizable gravity theory
with $z=3$ \cite{Horava:2009uw}, i.e.~including a ${\bf k}^6$ term. Note that in our case the $z=2$ contribution to the propagator
is a result of our NC deformation of the Euclidean space. The NC scale $\lambda$ and the slope of $\vartheta(x^0)$
set the scale of the propagator modification, which we assume to be the Planck scale. At small momenta our propagator can be
 approximated by the standard, $\mathbb{R}^d$ isotropic, propagator $E^2+{\bf k}^2+m^2$, leading to a theory which is
 approximately invariant under the Euclidean group $SO(d)\ltimes\mathbb{R}^d$ (up to $1/M_\mathrm{pl}$-corrections). 
 We show that the approximate $SO(d)\ltimes \mathbb{R}^d$-symmetry in the IR is still present at the quantum (one-loop) level.

\subsection{One-loop structure of our NC $\Phi^4$-theory}
We now focus on an interacting QFT with the potential $V[\Phi]=\frac{1}{2}m^2\Phi^2 +\frac{g_4}{4!}\Phi^4$ 
and $d=4$ in (\ref{eqn:action}), with the choice of deformation as described in Subsection \ref{par:z2}. 
The inverse propagator of this theory is given by (\ref{eqn:prop}).
 As argued above, the $\star$-products drop out of the potential, thus we do not obtain UV/IR-mixing
when calculating loop amplitudes. Moreover, due to the additional ${\bf k}^4$-part in the propagator denominator, the divergences
are tamed and we find for the one-loop correction $\Pi$ to the two-point function
\begin{flalign}
\label{eqn:divergence}
\Pi\big\vert_{\Lambda_E/\Lambda_k=\text{const}} =-\frac{g_4}{\sqrt{8}\,\pi^2}\sqrt{\frac{\Lambda_E}{\alpha^3}} +\text{finite}~,
\end{flalign}
where $g_4$ denotes the coupling constant and $\Lambda_{E}$ and $\Lambda_k$ denote 
the momentum space cutoff for $E$ and $k=\Vert{\bf k}\Vert$,
respectively.  
The ratio $\Lambda_E/\Lambda_k$ is kept fixed for the limit $\Lambda_E,\Lambda_k\to\infty$. 
The one-loop corrections to the four-point function are found to be finite.
Thus, we obtain that our NC deformation improves the quantum behaviour of perturbative $\Phi^4$-theory,
without introducing the UV/IR-mixing, which is typical for other NC QFTs. 
On top of that, the Landau pole, which typically arises through the renormalization group running of the 
four-point coupling $g_4$ in commutative $\Phi^4$-theory, is absent
in our NC theory at the one-loop level due to the finiteness of the four-point function.
Power-counting suggests that this result extends to higher orders in the perturbation theory.
The result (\ref{eqn:divergence}) shows that the $SO(4)$-violating ${\bf k}^4$-term receives no renormalization at the one-loop level,
 thus the IR limit of the loop-corrected propagator is again of the form $E^2+{\bf k}^2+m^2$ plus Planck suppressed corrections.
This leads to a phenomenologically acceptable theory in the IR.

\section{\label{sec:conclusion}Conclusions and outlook}
We have investigated a class of NC deformations of the Euclidean space leading to an unmodified algebra 
of smooth functions, but a deformed differential calculus. In contrast to the standard Moyal-Weyl case, 
this yields a scalar field theory with modified kinetic term and undeformed local potential.
 For a particular choice of deformation we have obtained a $z=2$ anisotropic 
scalar field propagator, leading to
improved properties of the four-dimensional deformed $\Phi^4$-theory at the quantum level. 
In particular, the one-loop corrections to the four-point coupling are finite, thus avoiding the Landau pole. 
Since for our NC deformation the UV/IR-mixing is absent, the propagator
reduces to the standard $SO(4)$ isotropic propagator in the IR limit, with Planck scale suppressed symmetry violating 
corrections.
Our theory illustrates that NC geometry is a rich framework, which can lead
to effects very different from the phase factors in the interactions as obtained on Moyal-Weyl deformed Euclidean space.

Ultimately, it would be interesting to apply a similar deformation to NC gravity \cite{Aschieri} with the hope
of constructing a perturbatively renormalizable, or at least better behaved, gravity theory on NC spaces.

\section*{Acknowledgements}
We thank Harald Grosse and Thorsten Ohl for valuable comments and discussions.
CFU is supported by the German National Academic Foundation 
(Studienstiftung des deutschen Volkes).
AS and CFU are supported by Deutsche
Forschungsgemeinschaft through the Research Training Group GRK\,1147 
\textit{Theoretical Astrophysics and Particle Physics}.


\begin{thebibliography}{10}
\bibitem{Grosse}
  H.~Grosse and R.~Wulkenhaar,
  Commun.\ Math.\ Phys.\  {\bf 254}, 91 (2005);
  H.~Grosse and R.~Wulkenhaar,
  Commun.\ Math.\ Phys.\  {\bf 256}, 305 (2005);
  H.~Grosse and R.~Wulkenhaar,
  Eur.\ Phys.\ J.\  C {\bf 35}, 277 (2004);
  H.~Grosse and R.~Wulkenhaar,
  arXiv:0909.1389 [hep-th].

\bibitem{Rivasseau}
  V.~Rivasseau, F.~Vignes-Tourneret and R.~Wulkenhaar,
  Commun.\ Math.\ Phys.\  {\bf 262}, 565 (2006);
  R.~Gurau, J.~Magnen, V.~Rivasseau and F.~Vignes-Tourneret,
  Commun.\ Math.\ Phys.\  {\bf 267}, 515 (2006);
  M.~Disertori, R.~Gurau, J.~Magnen and V.~Rivasseau,
  Phys.\ Lett.\  B {\bf 649}, 95 (2007).

\bibitem{RGE}
  R.~Gurau and O.~J.~Rosten,
  JHEP {\bf 0907}, 064 (2009);
  A.~Sfondrini and T.~A.~Koslowski,
  arXiv:1006.5145 [hep-th].


\bibitem{Fischer}
  A.~Fischer and R.~J.~Szabo,
  JHEP {\bf 0902}, 031 (2009);
  A.~Fischer and R.~J.~Szabo,
  arXiv:1001.3776 [hep-th].
  
 \bibitem{Grosse:2005iz}
  H.~Grosse and M.~Wohlgenannt,
  Nucl.\ Phys.\  B {\bf 748}, 473 (2006).
  
  \bibitem{fuzzyUVIR}
  C.~S.~Chu, J.~Madore and H.~Steinacker,
  JHEP {\bf 0108}, 038 (2001);
  H.~Steinacker,
  JHEP {\bf 0503}, 075 (2005);
  M.~Panero,
  JHEP {\bf 0705}, 082 (2007).
  
  \bibitem{Horava:2009uw}
  P.~Horava,
  Phys.\ Rev.\  D {\bf 79}, 084008 (2009).
  
  \bibitem{Reshetikhin:1990ep}
  N.~Reshetikhin,
  Lett.{} Math.{} Phys.{}  \textbf{20}, 331 (1990).

\bibitem{Jambor:2004kc}
  C.~Jambor and A.~Sykora,
  [arXiv:hep-th/0405268].
  
  \bibitem{Aschieri}
  P.~Aschieri, C.~Blohmann, M.~Dimitrijevic, F.~Meyer, P.~Schupp and J.~Wess,
  Class.\ Quant.\ Grav.\  {\bf 22}, 3511 (2005);
  P.~Aschieri, M.~Dimitrijevic, F.~Meyer and J.~Wess,
  Class.\ Quant.\ Grav.\  {\bf 23}, 1883 (2006).
  
  \bibitem{Ohl:2009qe}
  T.~Ohl and A.~Schenkel,
  arXiv:0912.2252 [hep-th].


\end{thebibliography}
\end{document}